\documentclass[review]{elsarticle}

\usepackage{lineno,hyperref}
\usepackage{pdflscape}
\usepackage{amsmath, amsthm, amssymb, amsfonts}
\modulolinenumbers[5]

\journal{Nuclear Physics A}

\begin{document}

\begin{frontmatter}

\title{
Flavor-dependent eigenvolume interactions \\
in a hadron resonance gas
}

\author[add1]{P. Alba}
\ead{alba@fias.uni-frankfurt.de}
\author[add1,add2,add3]{V. Vovchenko}
\author[add1,add4]{M. I. Gorenstein}
\author[add1,add2,add5]{H. Stoecker}

\address[add1]{Frankfurt Institute for Advanced Studies, Goethe Universit\"at Frankfurt,
D-60438 Frankfurt am Main, Germany}
\address[add2]{Institut f\"ur Theoretische Physik,
Goethe Universit\"at Frankfurt, D-60438 Frankfurt am Main, Germany}
\address[add3]{Department of Physics, Taras Shevchenko National University of Kiev, 03022 Kiev, Ukraine}
\address[add4]{Bogolyubov Institute for Theoretical Physics, 03680 Kiev, Ukraine}
\address[add5]{GSI Helmholtzzentrum f\"ur Schwerionenforschung GmbH, D-64291 Darmstadt, Germany}

\begin{abstract}
Eigenvolume effects in the hadron resonance gas (HRG) model are studied for experimental hadronic yields in nucleus-nucleus collisions. If particle eigenvolumes are different for different hadron species, the excluded volume HRG (EV-HRG) improves fits to multiplicity data. In particular, using different mass~-~volume relations for strange and non-strange hadrons we observe a remarkable improvement in the quality of the fits. This effect appears to be rather insensitive to other details in the schemes employed in the EV-HRG. We show that the parameters found from fitting the data of the ALICE Collaboration in central Pb+Pb collisions at the collision energy $\sqrt{s_{\rm NN}} = 2.76$~TeV entail the same improvement for all centralities at the same collision energy, and for the RHIC and SPS data at lower collision energies. Our findings are put in the context of recent fits of lattice QCD results.
\end{abstract}

\end{frontmatter}

\section{Introduction}
\label{intro}

The statistical model approach is an important tool to extract the properties of matter created in relativistic nucleus-nucleus (A+A) collisions (see, e.g., Refs.~\cite{Stoecker:1981za,Cleymans:1992zc,Yen:1998pa,Becattini:2000jw,BraunMunzinger:2001ip,Rafelski:2002ga,Andronic:2005yp,Vovchenko:2015idt}). The experimental hadron multiplicities in A+A collisions have been successfully fitted in a wide range of collision energies in terms of a few basic parameters: temperature $T$, baryon chemical potential $\mu_B$, and system volume $V$. One popular version of the statistical model is the {\it ideal} (point-particle) hadron resonance gas (I-HRG). It is based on the idea that the resonance formation mediates the attractive interactions among hadrons. This corresponds to a system of non-interacting hadrons and resonances, hence the following formula for the system pressure in the grand canonical ensemble:
\begin{equation}\label{id-press}
 p_{\rm I}(T,\mu_B)~=~\sum_j\,p_j^{\rm id}(T,\mu_j)~,
\end{equation}
where $p_j^{\rm id}$ is the ideal gas pressures, and the sum runs over all known hadrons and resonances. The chemical potentials $\mu_j$ for $j$th particle species are taken as:
\begin{equation}
\mu_j~=~b_j\mu_B~+~s_j\mu_S~+~q_j\mu_Q~,
\end{equation}
where $b_j$, $s_j$, and $q_j$ correspond, respectively, to the baryon number, strangeness, and electric charge of $j$th particle. The pressure function (\ref{id-press}) depends on two independent variables: temperature $T$ and baryon chemical potential $\mu_B$. The strange chemical potential $\mu_S=\mu_S(T,\mu_B)$ and the electric chemical potential $\mu_Q=\mu_Q(T,\mu_B)$ are found from the conditions of zero net-strangeness and fixed electric-to-baryon charge ratio in the colliding nuclei. Other intensive observables (e.g. particle number densities, energy density, and entropy density) are obtained from (\ref{id-press}) by standard thermodynamic relations.

\section{Repulsive interactions}
\label{sec:1}

In order to account for the short-range repulsive interactions
between hadrons a thermodynamically consistent {\it excluded volume} (EV)
van der Waals procedure was suggested in \cite{Rischke:1991ke}. For the multi-component HRG
the simplest formulation of the EV model leads to a transcendental equation~\cite{Yen:1998pa}:
\begin{eqnarray}
 \label{p}
 & p(T,\mu_B)~=~\sum_j\,p_j^{\rm id}(T,\mu_j^*)~,\\
 & \mu_j^*~=~\mu_j~-~v_j~p(T,\mu_B)~.\label{mu}
\end{eqnarray}
with $v_j$
%=16\pi\,r_j^3/3$
being the eigenvolume parameter for the particle $j$.
%, and $r_j$ corresponds to its effective hard-core radius.
The particle number densities of $i$th species are then calculated as:
\begin{equation}
\label{ni}
n_i(T,\mu_B)~=~\frac{n_i^{\rm id}(T,\mu_i^*)}{1+\sum_j\,v_j\,n_j^{\rm id}(T,\mu_j^*)}\,.
\end{equation}
The I-HRG is usually considered as a baseline with regards to both the fit of the data on hadron multiplicities \cite{BraunMunzinger:1994xr,Alba:2014eba} and the comparison with lattice QCD simulations at $\mu_B=0$ (see, e.g., Refs.~\cite{Borsanyi:2013bia,Bazavov:2014pvz,Braun-Munzinger:2014lba}). 
At the same time, we think that repulsive interactions between hadrons due to non-zero size of particles are both physically justified and important (see, e.g., Refs.~\cite{Andronic:2012ut,Begun:2012rf}). Some particular implementations of the EV-HRG model have been confronted to the lattice QCD data~\cite{Andronic:2012ut,Vovchenko:2014pka,Albright:2014gva}. In the present paper we discuss the fit of mean hadron multiplicities and the role of EV effects. In \cite{Alba:2017bbr} has been performed a systematic study of the same EV schemes here presented together with a modified hadronic list with undetected higher-mass resonances, and it has been shown how their combined effect leads to a consistent description of lattice QCD thermodynamics and experimental measurements of hadronic yields.\\

Equation (\ref{ni}) gives the {\it primordial} equilibrium density of stable hadrons and resonances in A+A collisions. Their total numbers $N_i$ are obtained multiplying $n_i$ by the system volume $V$. The {\it final} multiplicity $\langle N_h\rangle$ is calculated in the HRG model as a sum of the primordial multiplicity and resonance decay contributions as follows:
\begin{equation}\label{Nh}
\langle N_h \rangle ~=~V\,n_h~+~V\,\sum_R\langle n_h\rangle_R\,n_R~,
\end{equation}
where $\langle n_h\rangle_R$ is the average number of particles of type $h$ which result from decay of a resonance $R$. As seen from Eq.~(\ref{ni}) the EV procedure introduces a suppression factor  $[1+\sum_j\,v_j\,n_j^{id}(T,\mu_j^*)]^{-1} <1$, which is the same for all types of particles, and an additional suppression due to the shift of chemical potential as given by Eq.~(\ref{mu}), which in the classical (Boltzmann) approximation leads to the factor $\exp[-\,v_ip(T,\mu_B)]< 1$. When $v_i$ is the same for all particle species, the overall suppression of each hadron density $n_i$ due to the EV effects compared to their ideal gas values $n_i^{\rm id}$ is essentially the same, with small differences resulting from quantum-statistical effects. Therefore the particle number ratios are almost unchanged, and rescaling the total volume $V$ one obtains $\langle N_h\rangle$ values equal to those in the I-HRG. As a consequence, the $v_i={\rm const}$ case yields essentially no changes in the behaviour of the thermal fits to hadron yield data\footnote{The constant eigenvolume for all hadron species is still relevant for the description of lattice QCD, as it influences the HRG equation of state and fluctuations of the conserved charges \cite{Alba:2017bbr}.}. However, if $v_i$ are chosen to be different for different $i$, the suppression will be stronger for particles with a larger eigenvolume. In such a case the description of the hadron ratios is affected notably (see e.g. Refs.~\cite{Yen:1998pa,Bugaev:2012wp,Vovchenko:2017zpj,Vovchenko:2015cbk}).

\section{Analysis of experimental data}

\subsection{Central Pb+Pb collisions at ALICE}

We employ the EV-HRG model (\ref{p})-(\ref{ni}) to describe experimental data on particle yields at mid-rapidity for $\pi^{\pm}$, $K^{\pm}$, $p$($\bar p$) \cite{Abelev:2013vea}, $K^0_S$, $\Lambda$ \cite{Abelev:2013xaa}, $\Xi^{\mp}$, $\Omega +\bar\Omega$ \cite{ABELEV:2013zaa,Becattini:2014hla} and $\phi$ \cite{Abelev:2014uua} in Pb+Pb collisions at $\sqrt{s_{NN}}$=2.76~TeV in the 0~-~5~\% centrality class measured by the ALICE Collaboration at the Large Hadron Collider (LHC) of European Organization for Nuclear Research (CERN). In our calculations we include hadrons and resonances that are listed by the Particle Data Group \cite{Patrignani:2016xqp}; we employ here the list called PDG2014  in Ref.~\cite{Alba:2017bbr}. For more details on the implementation of the model see \cite{Alba:2014eba,Nahrgang:2014fza,Bluhm:2014wha}. Note that we do not include light nuclei, neither in the fit nor in the particle list. A thermal fit for the light nuclei has been considered in Refs.~\cite{Adam:2015vda,Adam:2015yta,Andronic:2017pug}, however it is still highly debated whether the production mechanism for nuclei is thermal or should rather be attributed to the coalescence processes. The HRG model fits are performed by minimising the value of:
\begin{equation}\label{xi}
\frac{\chi ^2}{N_{\rm dof}}~=\frac{1}{N_{\rm dof}}~\sum_{h} \frac{\left( \langle N_h^{\rm exp}\rangle~-~
\langle N_h \rangle \right)^2}{\sigma_h^2}~,
\end{equation}
where $\langle N_h^{\rm exp}\rangle $ and $\langle N_h\rangle $ are respectively the experimental and HRG model hadron yields, $N_{\rm dof}$ is the number of degrees of freedom, i.e. the number of data points minus the number of fitting parameters, and $\sigma_h$ are experimental errors on hadron yields.
The bag-like parametrisation for the particle eigenvolumes was proposed in Refs.~\cite{Vovchenko:2015cbk} and \cite{Vovchenko:2016ebv}; this implies the linear mass-volume relation ($m_i$ is a mass of $i$th particle):
\begin{equation}
\label{direct_mass}
v_i~=~\alpha ~{m_i}~,
\end{equation}
with the same parameter $\alpha$ for all particle species. As an exotic alternative of the bag-model relation (\ref{direct_mass}) the inverse mass-volume relation:
\begin{equation}
\label{inverse_mass}
v_i~=~\gamma~m_i^{-1}~,
\end{equation}
will be also considered in the present study for strange hadrons. In general, the current experimental measurements on
hadronic ground states \cite{Patrignani:2016xqp} do not allow to select any specific trend for their sizes, and excited states can either be larger, due to radial and angular excitations in the quark model framework, or smaller if other exotic degrees of freedom \cite{Friedmann:2009mx,Friedmann:2009mz,Friedmann:2012kr} are assumed.

The following EV-HRG schemes will be considered in our investigation:
\begin{description}
\item[1b] Eq.~(\ref{direct_mass}) is assumed for all hadrons, with a common $\alpha$ parameter.
\item[2b] Eq.~(\ref{direct_mass}) is assumed for all hadrons, but with different values of the $\alpha$ parameter
for non-strange and strange hadrons.
\item[4b] Eq.~(\ref{direct_mass}) is assumed for all hadrons
with different values of the $\alpha$ for non-strange mesons, strange mesons, non-strange
baryons, and strange baryons.
\item[s-inv] Eq.~(\ref{direct_mass}) is assumed for non-strange hadrons while Eq.~(\ref{inverse_mass}) applies to strange ones.
\end{description}
Note that the same schemes have already been studied for pure gauge theories \cite{Alba:2016fku} and lattice QCD simulations \cite{Alba:2017bbr}.

The classical EV relation $v_i=16\pi r_i^3/3$ connects the particle eigenvolume parameter $v_i$ to its effective radius parameter $r_i$.
We will use the latter in the present study for the sake of simplicity.
The parameters $\alpha$ and/or $\gamma$
will be fixed by specifying the radius parameter value of a single ground-state hadron representative of the hadronic family under investigation.

Table \ref{tab-alice05-schemes} lists the results of the fits with different EV schemes. In the 1b and 2b schemes the proton radius $r_p$ is not included in the fitting procedure, but rather fixed to the values obtained from the fit to lattice QCD thermodynamics with the same hadron list \cite{Alba:2017bbr}. This is done to avoid the instability of thermal fits for experimental data in these schemes, related to the appearance of the "second minimum" structure in the $\chi^2$ as a function of $T$ which manifests at very high temperatures~\cite{Vovchenko:2015cbk,Vovchenko:2016ebv}; it should be noted that the second minimum generally corresponds to very large values of the packing factor, stressing the condition of a dilute system where the EV-HRG model is more reliable. The same instabilities appear in the 4b scheme, for which however it is possible to locate a local minimum at low temperature without any constraint from lattice. Due to this choice the $r_\Lambda$ in the 4b scheme is affected by a very large uncertainty, which could be reduced to the same magnitude as for the other parameters through a detailed investigation of the $\chi^2$ profile. These instabilities could be indeed a new feature of hadronic matter connected with systematic large uncertainties in particle yields \cite{Vovchenko:2015cbk}, but we leave this discussion to other studies.

\begin{figure}[h]
\begin{center}
\includegraphics[width=0.84\textwidth]{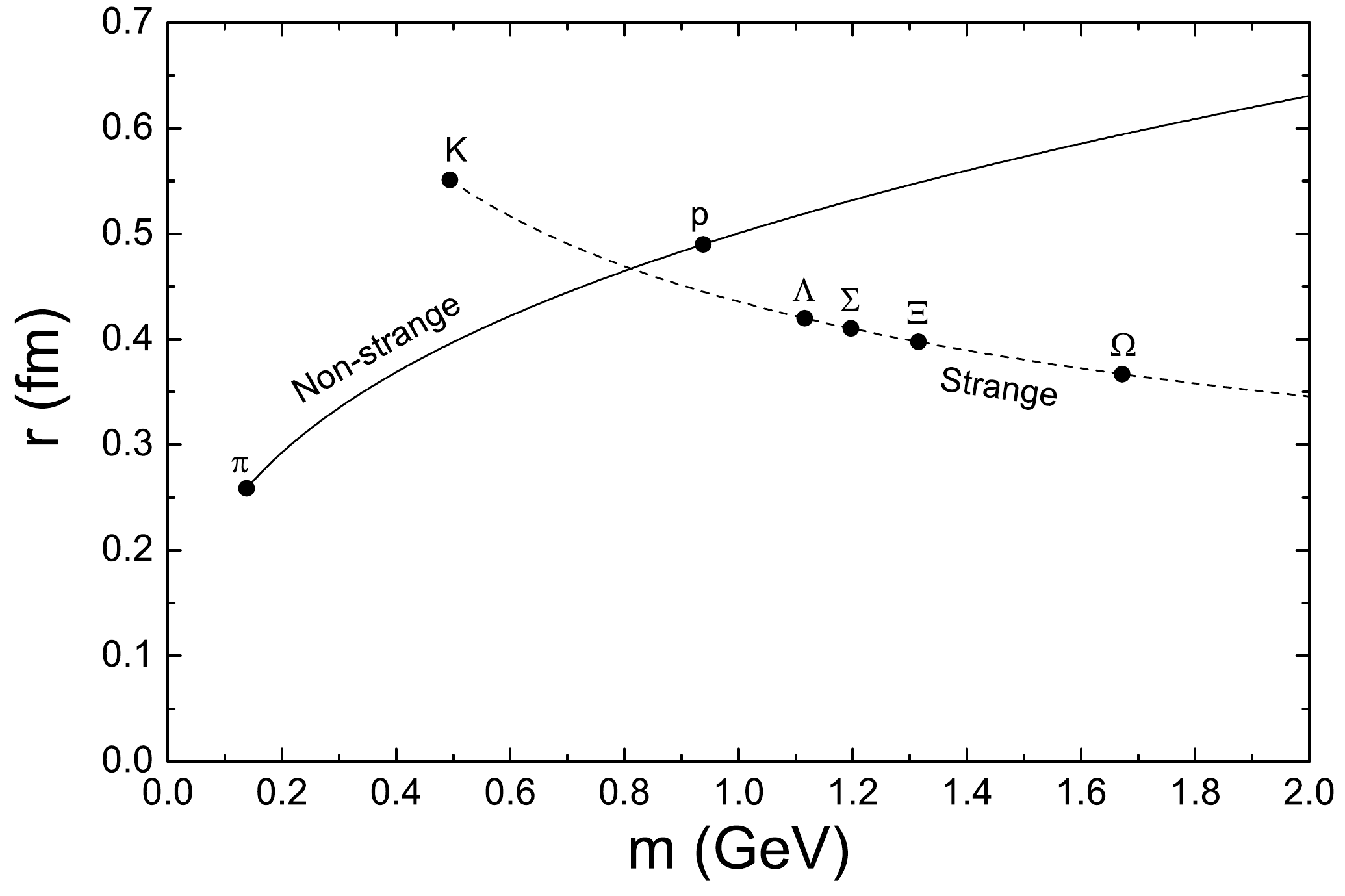}
\caption{The effective radii parameters of non-strange (solid line) and strange (dashed line) hadrons given respectively by the parameters extracted from ALICE data for the s-inv scheme (Table \ref{tab-alice05-schemes}).}
\label{radiuses}
\end{center}
\end{figure}

The small value of $r_p$ in the 1b scheme leads to negligible effects on the fit of the hadron yields, and indeed in this case the results are very close to the I~-~HRG ones. In the 2b scheme there is a notable improvement, which is related to the different EV interactions for strange and non-strange hadrons with the introduction of an additional free parameter $r_\Lambda$ for all strange particles; here we have the non trivial result of smaller strange states with respect to light ones with equal mass ($r_\Lambda < r_p$), which is compatible with what is found from fits to lattice QCD \cite{Alba:2017bbr}.
With respect to the 2b scheme, the 4b one introduces a meson-baryon difference with 4 free parameters: $r_\pi$, $r_K$, $r_p$, and $r_\Lambda$; this leads to an even more essential improvement of the quality of the fit. It is interesting that in the 4b scheme $r_\pi < r_p$ and $r_K > r_\Lambda$ (central values), which is to some extent compatible with the dependencies considered in the s-inv scheme. Indeed it is with the s-inv scheme that we obtain the best description of the ALICE data in terms of the reduced $\chi^2$; the behaviour of the radius parameters as functions of hadron masses in the s-inv scheme is shown in Fig.~\ref{radiuses}.

\begin{landscape}

\begin{table}[h]
\begin{center}
\begin{tabular}{lcccccc}
\hline\noalign{\smallskip}
 & $\chi^2/N_{\rm dof}$ &  T (MeV) & $\mu_B$ (MeV) & V (fm$^3$) & r$_i$ (fm)\\
\noalign{\smallskip}\hline\noalign{\smallskip}
I-HRG & 22.2/9 $\simeq$ 2.4 &  154.1$\pm$2.2 & 3.4$\pm$6.9 & 4993$\pm$644 & - \\
\noalign{\smallskip}\hline\noalign{\smallskip}
1b & 22.2/9 $\simeq$ 2.4 & 154.5$\pm$2.2 & 3.5$\pm$6.9 & 5009$\pm$643 & r$_p$=0.13\\
\noalign{\smallskip}\hline\noalign{\smallskip}
2b & 12.5/8 $\simeq$ 1.5 & 159.6$\pm$3.6 & 2.4$\pm$7.7 & 5540$\pm$647 & \small \begin{tabular}{@{}c@{}}r$_p$=0.374 \\ r$_\Lambda$=0.313$\pm$0.042\end{tabular}\\
\noalign{\smallskip}\hline\noalign{\smallskip}
4b & 0.86/5 $\simeq$ 0.172 & 150.8$\pm$5.4 & 2.3$\pm$7.7 & 8303$\pm$2522 & \footnotesize \begin{tabular}{@{}cc@{}}r$_p$=0.402$\pm$0.081 & r$_\pi$=0.179$\pm$0.083 \\ r$_\Lambda$=0.106$\pm$1.467 & r$_K$=0.361$\pm$0.08\end{tabular}\\
\noalign{\smallskip}\hline
s-inv & 0.88/7 $\simeq$ 0.098 & 152.9$\pm$1.8 & 2.6$\pm$7.7 & 9091$\pm$713 & \small \begin{tabular}{@{}c@{}}r$_p$=0.449$\pm$0.044 \\ r$_\Lambda$=0.371$\pm$0.044\end{tabular}\\
\noalign{\smallskip}\hline\noalign{\smallskip}
\end{tabular}
\end{center}
\caption{Parameters of fits to the ALICE data in Pb+Pb collisions at 0-5\% centrality with different schemes of the HRG model.}
\label{tab-alice05-schemes}
\end{table}

\end{landscape}

Note that the parameters extracted here from hadron yields are compatible with the ones found in \cite{Alba:2017bbr} from EV-HRG fits to lattice data, where it is pointed out how small values of $r_\Lambda$ consistent with zero can be a direct consequence of missing strange baryons, whose relevance has already been established in the literature \cite{Bazavov:2014xya,Alba:2017mqu}.\\
The results presented in Table~\ref{tab-alice05-schemes} confirm the strong influence of the flavour dependent EV interactions on the fit to the measured hadron yields. On the other hand, the freeze-out temperature values listed in Table \ref{tab-alice05-schemes} do not vary strongly and are rather consistent with the fit results of various other groups~\cite{Becattini:2014hla,Petran:2013lja,Stachel:2013zma,Floris:2014pta,Andronic:2017pug}. The finite size of hadrons leads to an increment of the system volume $V$ in the EV-HRG with respect to the I-HRG; thus the densities characterising the thermal fireball at the chemical freeze-out are smaller, e.g. typical values of the energy density in the s-inv scheme are 0.15-0.20 GeV/fm$^3$, lower than the 0.25-0.40 GeV/fm$^3$ obtained from the I-HRG. However, the total particle density is decreasing accordingly, and energy per particle is a quite robust quantity with $E/N\cong 0.83$ GeV for the s-inv EV-HRG and $E/N\cong 0.91$ GeV for the I-HRG.\\

\begin{figure}[h]
\begin{center}
\includegraphics[width=0.79\textwidth]{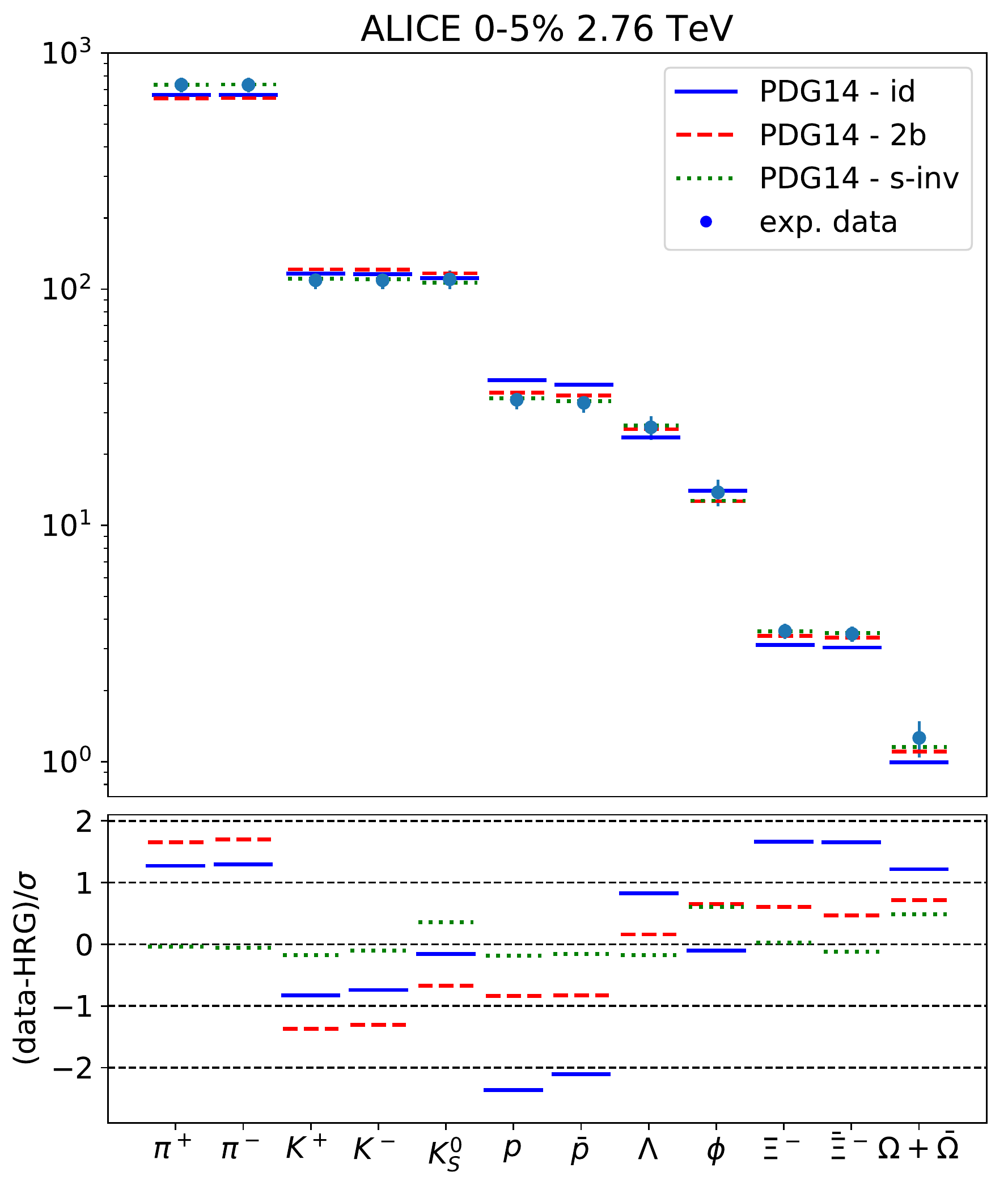}
\caption{Fit to ALICE data with respectively ideal (blue solid), 2b (red dashed) and s-inv (green dotted) schemes of the HRG model.}
\label{fit-yields-alice}
\end{center}
\end{figure}

In Fig. \ref{fit-yields-alice} we show the ALICE yields in comparison with the model calculations within I-HRG and EV-HRG with 2b and s-inv schemes with the corresponding parameters listed in Table \ref{tab-alice05-schemes}. Compared to I-HRG, the EV effects within both 2b and s-inv schemes simultaneously lead to a suppression of (anti-)proton yields and to the relative enhancement of strange baryons yields. Both these findings are in agreement with the study of lattice data performed in Ref.~\cite{Alba:2017bbr}.
Note that the ``anomaly'' of the $p/\pi$ ratio, which is poorly described by the I-HRG and which has sparked some discussion~\cite{Stachel:2013zma,Floris:2014pta,Steinheimer:2012rd,Becattini:2012xb}, here is described very well solely due to the eigenvolume interactions. We do point out that different mechanisms, which were proposed to explain the $p/\pi$ anomaly, could be considered together with the EV effects in a more involved model in order to test their relative relevance.

\begin{table}[h]
\begin{center}
\begin{tabular}{lllll}
\hline\noalign{\smallskip}
 & $\chi^2/N_{\rm dof}$ & T (MeV) & $\mu_B$ (MeV) & V (fm$^3$)\\
\noalign{\smallskip}\hline\noalign{\smallskip}
ALICE 5-10\% & 19.7/7 $\simeq$ 2.82 & 150.6$\pm$2.2 & -0.8$\pm$6.5 & 4793$\pm$613\\
ALICE 10-20\% & 45.6/9 $\simeq$ 5.07 & 158.5$\pm$2.1 & -0.4$\pm$6.0 & 2535$\pm$308\\
ALICE 20-30\% & 50.1/8 $\simeq$ 6.26 & 158.2$\pm$2.3 & 1.2$\pm$6.2 & 1769$\pm$233\\
ALICE 30-40\% & 44.1/8 $\simeq$ 5.51 & 158.2$\pm$2.3 & 0.4$\pm$6.6 & 1148$\pm$147\\
ALICE 40-50\% & 35.2/8 $\simeq$ 4.4 & 158.8$\pm$2.3 & 0.3$\pm$6.5 & 688$\pm$91.0\\
ALICE 50-60\% & 16.3/8 $\simeq$ 2.04 & 155.3$\pm$2.4 & -2.5$\pm$7.5 & 463$\pm$62.5\\
ALICE 60-70\% & 12.8/8 $\simeq$ 1.61 & 150.8$\pm$3.0 & -1.7$\pm$10.9 & 296$\pm$46.7\\
ALICE 70-80\% & 13.8/8 $\simeq$ 1.73 & 152.8$\pm$3.5 & -1.9$\pm$10.5 & 122.2$\pm$22.4\\
\noalign{\smallskip}\hline
\end{tabular}
\end{center}
\caption{Fits of the ALICE data at different centralities with I-HRG.}
\label{tab-central-ideal}
\end{table}

\begin{table}[h]
\begin{center}
\begin{tabular}{lllll}
\hline\noalign{\smallskip}
 & $\chi^2/N_{\rm dof}$ & T (MeV) & $\mu_B$ (MeV) & V (fm$^3$)\\
\noalign{\smallskip}\hline\noalign{\smallskip}
ALICE 5-10\% & 1.022/7 $\simeq$ 0.14 & 154.3$\pm$2.3 & 0.019$\pm$6.9 & 7141$\pm$633\\
ALICE 10-20\% & 2.7/9 $\simeq$ 0.30 & 156.7$\pm$1.6 & -1.8$\pm$6.9 & 5065$\pm$355\\
ALICE 20-30\% & 6.08/8 $\simeq$ 0.76 & 158.4$\pm$1.8 & -2.9$\pm$7.3 & 3269$\pm$249\\
ALICE 30-40\% & 6.9/8 $\simeq$ 0.86 & 158.7$\pm$1.9 & -2.5$\pm$7.6 & 2141$\pm$155\\
ALICE 40-50\% & 3.07/8 $\simeq$ 0.38 & 158.0$\pm$1.8 & -3.0$\pm$7.7 & 1333$\pm$99.5\\
ALICE 50-60\% & 4.42/8 $\simeq$ 0.55 & 155.3$\pm$2.0 & -2.1$\pm$8.8 & 823$\pm$65.3\\
ALICE 60-70\% & 8.09/8 $\simeq$ 1.01 & 153.2$\pm$2.9 & -3.1$\pm$12.5 & 448$\pm$44.2\\
ALICE 70-80\% & 5.01/8 $\simeq$ 0.62 & 161.2$\pm$4.5 & -1.9$\pm$11.7 & 164.1$\pm$21.7\\
\noalign{\smallskip}\hline
\end{tabular}
\end{center}
\caption{Fits of the ALICE data at different centralities with EV-HRG in the s-inv scheme, where $r_p$ and $r_\Lambda$ are fixed to the values listed in Table \ref{tab-alice05-schemes}.}
\label{tab-central-sinv}
\end{table}

\subsection{Centrality dependence at ALICE}

Results of the fits to ALICE data at 2.76 TeV at different centralities are presented for the ideal and s-inv schemes in Tables \ref{tab-central-ideal} and \ref{tab-central-sinv}, respectively. The $r_p$ and $r_\Lambda$ parameters for the s-inv scheme are the same listed in Table~\ref{tab-alice05-schemes} for 0-5\% centrality. The improvement of the ALICE data description within the s-inv scheme systematically persists across all other available centrality classes, yielding $\chi^2/N_{\rm dof} \lesssim 1$. The introduction of EV effects leads to an increase of volumes for all centralities by an average factor of $V^{\rm s-inv}/V^{\rm id}= 1.79\pm0.09$. Baryon chemical potential is consistent with zero at all centralities as expected.

\begin{table}[h]
\begin{center}
\begin{tabular}{lllll}
\hline\noalign{\smallskip}
 & $\chi^2/N_{\rm dof}$ & T (MeV) & $\mu_B$ (MeV) & V (fm$^3$)\\
\noalign{\smallskip}\hline\noalign{\smallskip}
STAR 200 & 16.8/8 $\simeq$ 2.1 & 161.7$\pm$2.2 & 29.2$\pm$8.3 & 1708$\pm$214\\
NA49 158 & 70.1/10 $\simeq$ 7.01 & 151.6$\pm$2.2 & 280.0$\pm$5.5 & 3431$\pm$436\\
NA49 80 & 71.7/7 $\simeq$ 10.2 & 142.7$\pm$4.2 & 338.01$\pm$4.7 & 3771$\pm$868\\
NA49 40 & 44.2/8 $\simeq$ 5.5 & 141.3$\pm$2.4 & 413.02$\pm$5.7 & 2519$\pm$374\\
NA49 30 & 29.6/7 $\simeq$ 4.2 & 142.9$\pm$2.7 & 455.41$\pm$5.0 & 1770$\pm$274\\
NA49 20 & 29.7/6 $\simeq$ 4.9 & 112.6$\pm$4.1 & 498.2$\pm$2.8 & 6979$\pm$1856\\
\noalign{\smallskip}\hline
\end{tabular}
\end{center}
\caption{The energy scan with I-HRG.}
\label{tab-energ-ideal}
\end{table}

\begin{table}[h]
\begin{center}
\begin{tabular}{lllll}
\hline\noalign{\smallskip}
 & $\chi^2/N_{\rm dof}$ & T (MeV) & $\mu_B$ (MeV) & V (fm$^3$)\\
\noalign{\smallskip}\hline\noalign{\smallskip}
STAR 200 & 6.5/8 $\simeq$ 0.82 & 158.1$\pm$1.6 & 29.7$\pm$8.8 & 3552$\pm$259\\
NA49 158 & 57.4/10 $\simeq$ 5.7 & 146.7$\pm$1.6 & 292.03$\pm$6.03 & 7685$\pm$534\\
NA49 80 & 28.3/7 $\simeq$ 4.05 & 144.4$\pm$2.4 & 349.2$\pm$5.5 & 6091$\pm$560\\
NA49 40 & 15.1/8 $\simeq$ 1.8 & 138.9$\pm$1.7 & 424.9$\pm$6.3 & 4967$\pm$358\\
NA49 30 & 7.08/7 $\simeq$ 1.01 & 138.5$\pm$1.9 & 459.5$\pm$5.5 & 4241$\pm$302\\
NA49 20 & 20.8/6 $\simeq$ 3.4 & 125.6$\pm$5.4 & 505.1$\pm$7.2 & 5033$\pm$1140\\
\noalign{\smallskip}\hline
\end{tabular}
\end{center}
\caption{The energy scan with EV-HRG in the s-inv scheme, where $r_p$ and $r_\Lambda$ are fixed to the values listed in Table \ref{tab-alice05-schemes}.}
\label{tab-energ-sinv}
\end{table}

\subsection{Lower collision energies}

\begin{figure}[t]
\begin{center}
\includegraphics[width=0.33\textwidth]{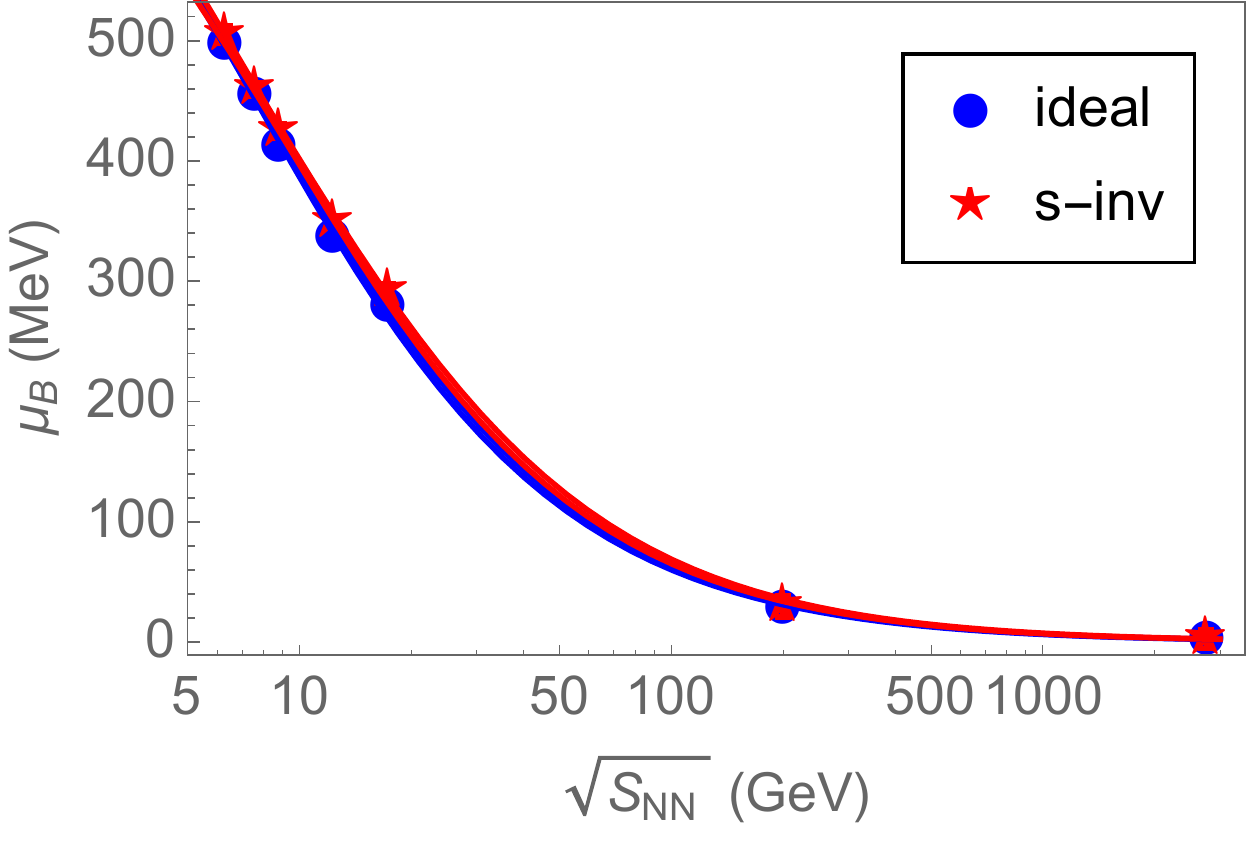}~~\includegraphics[width=0.33\textwidth]{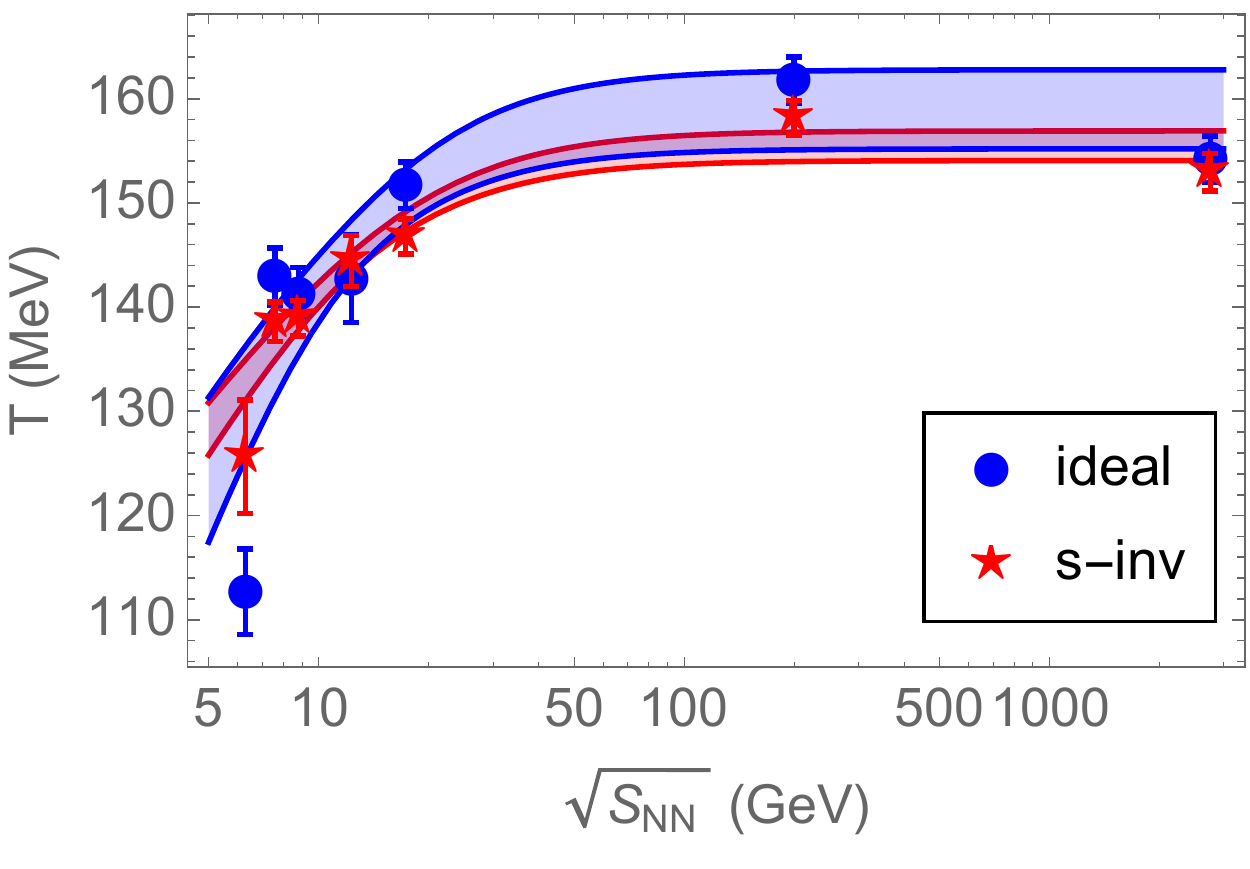}~~\includegraphics[width=0.33\textwidth]{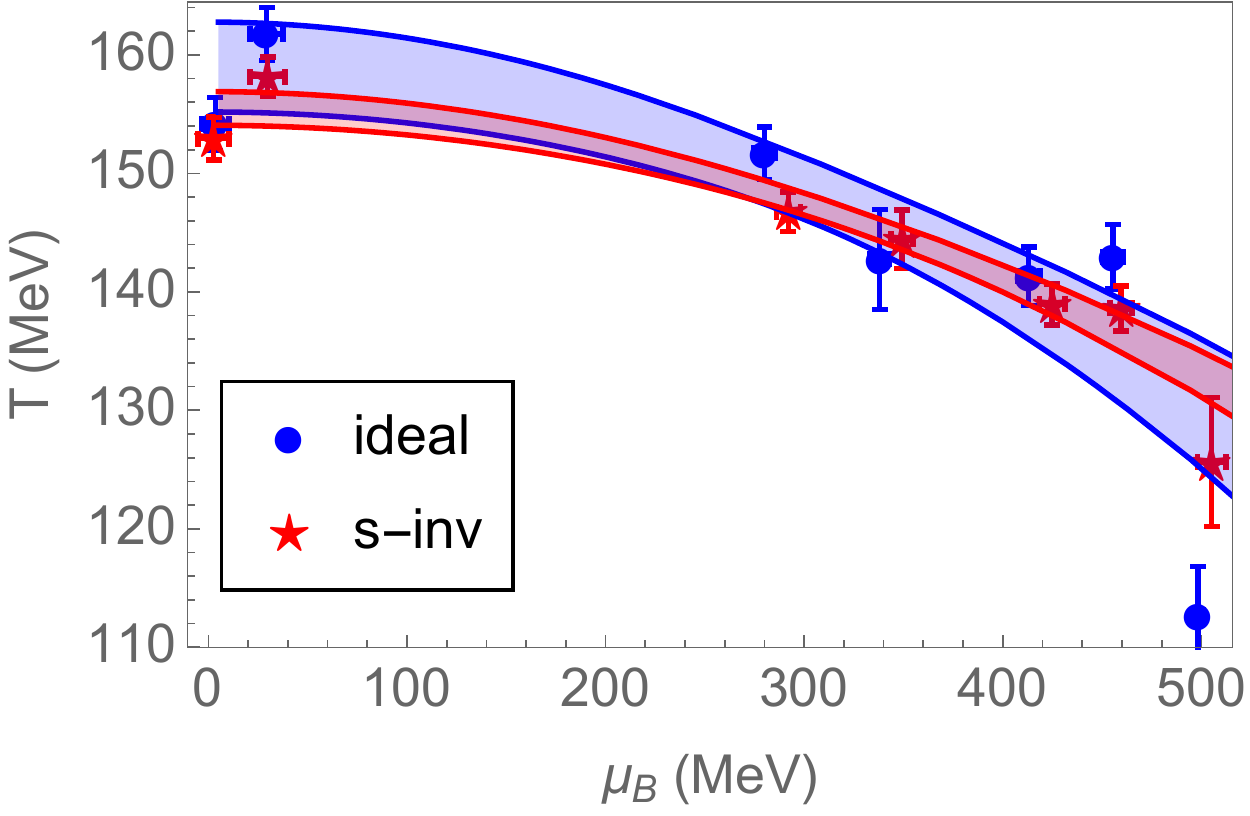}
\caption{Freeze-out values of $T$ and $\mu_B$ from Tables \ref{tab-alice05-schemes}, \ref{tab-energ-ideal} and \ref{tab-energ-sinv} for ideal HRG (blue dots) and EV-HRG in the s-inv scheme (red stars), as functions of collision energy (left and middle panels) and in the phase diagram (right panel). Corresponding interpolated bands from Eqs.~\ref{mub_energ} and \ref{t_phase} are shown on top of the data points.}
\label{freeze-out-energy}
\end{center}
\end{figure}

Results of the fits of STAR mid-rapidity data in Au+Au collisions at $\sqrt{s_{NN}}=200$~GeV \cite{Adams:2003xp,Adams:2006ke,Abelev:2007rw} and of NA49 $4\pi$ data in Pb+Pb collisions at different collision energies per nucleon $E_{\rm lab} = 20-158$~GeV \cite{Afanasiev:2002mx,Alt:2006dk,Alt:2007aa,Alt:2008qm,Alt:2008iv,Alt:2004kq,Anticic:2011ny,Friese:2002re} with the ideal and s-inv schemes are presented, respectively, in Tables \ref{tab-energ-ideal} and \ref{tab-energ-sinv}. The STAR and NA49 data correspond to 0-5\% (or 0-7\%) most central events.
The parameters $r_p$ and $r_\Lambda$ for the s-inv scheme are not fitted, but taken from the fit to the ALICE 0-5\% data~(Table~\ref{tab-alice05-schemes}). Without any specific tuning, the same EV parameters fixed to the ALICE data give a systematic improvement for lower energies, yielding in the s-inv scheme a $\chi^2/N_{\rm dof}$ reduced by a factor 2 or larger as compared to the ideal case.

Fig.~\ref{freeze-out-energy} depicts the extracted values for $T$ and $\mu_B$, within I-HRG and s-inv EV-HRG models, as functions of $\sqrt{s_{NN}}$ and in the $\mu_B$-$T$ plane. To interpolate the extracted parameters at different collision energies we use the following functions:
\begin{equation}
\label{mub_energ}
\mu_B=\frac{\mu_{0}}{1+\frac{\sqrt{s_{NN}}}{a}}~,~~~~~~T=T_{0}\left(1-\frac{b}{(a+\sqrt{s_{NN}})^2}\right)~;
\end{equation}
due to the very small uncertainties in the extracted $\mu_B$, instead of the coefficient $b$ we prefer to consider the following relation:
\begin{equation}
\label{t_phase}
T=T_{0}\left(1-\kappa_2\left(\frac{\mu_B}{T_0}\right)^2\right)~,
\end{equation}
from which $b=\kappa_2(\frac{\mu_0}{T_0}a)^2$. The resulting coefficients of the interpolations are listed in Table \ref{par-fo}, together with the $\chi^2/N_{\rm dof}=\chi^2_{phase}$ corresponding to Eq.~(\ref{t_phase}). While there is no specific improvement in the already well described $\mu_B$, the use of the s-inv schemes with respect to the ideal one, localise a narrower region in the phase diagram for the chemical freeze-out line, as also reflected in the smaller value of $\chi^2_{phase}$. The extracted $T_0^{s-inv}$ does not differ from the corresponding value at 2.76 TeV, and is fully compatible with the pseudo-critical temperature $T_c=154\pm9$ MeV extracted from lattice \cite{Bazavov:2011nk}.

\begin{figure}[t]
\begin{center}
\includegraphics[width=0.77\textwidth]{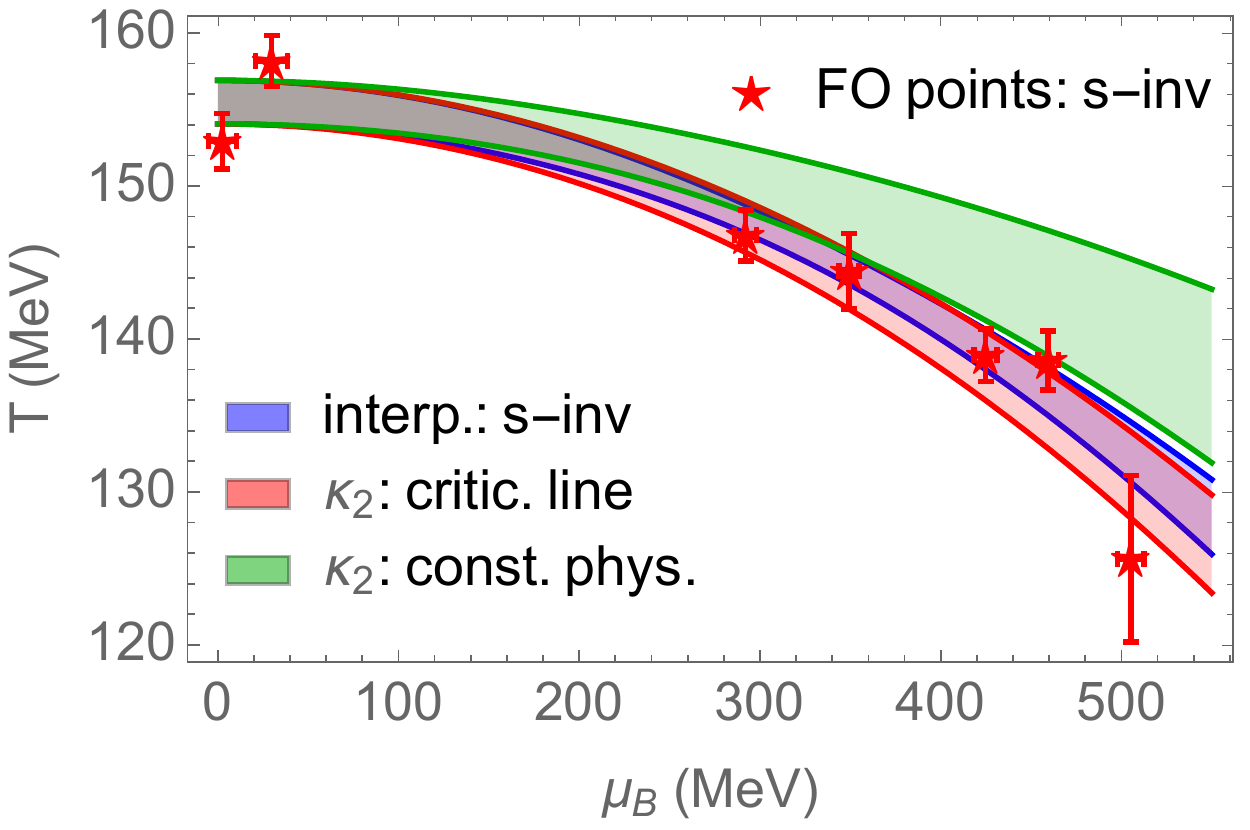}
\caption{The chemical freeze-out parameters extracted with the s-inv scheme in comparison with the corresponding interpolation with Eq.~(\ref{t_phase}), and with bands obtained employing curvatures extracted from lattice simulations \cite{Bellwied:2015rza,Bazavov:2017dus}.}
\label{freeze-out-phase-diagram}
\end{center}
\end{figure}

The values of $\kappa_2^{ideal}$ and $\kappa_2^{s-inv}$ are compatible, though $\kappa_2^{s-inv}$ has a noticeably smaller uncertainty and points to the lower end of the $\kappa_2^{ideal}$ error band. In Fig. \ref{freeze-out-phase-diagram} are shown the extracted freeze-out points within the s-inv scheme together with the corresponding interpolation, in comparison to the bands obtained by Eq.~(\ref{t_phase}) with $\kappa_2^{critic}=0.0148\pm0.0014$ extracted from the QCD pseudocritical line \cite{Bellwied:2015rza} and with $\kappa_2^{const}=0.0092\pm0.0028$ from lines of constant physics \cite{Bazavov:2017dus} both obtained from lattice QCD simulations; $T_0=T_0^{s-inv}$ is considered for all three cases.
The s-inv freeze-out band nicely falls within the pseudocritical one at all chemical potential values here considered, while the band corresponding to constant physics deviates at $\mu_B=$300 MeV. Note, however, that the $\mu_B > 300$~MeV region is already outside the range of applicability claimed in \cite{Bazavov:2017dus}.\\
The inclusion of 4th order terms in Eq.~(\ref{t_phase}) does not change the analysis of the s-inv freeze-out points, in contrast to \cite{Becattini:2016xct}, yielding small values for $\kappa_4$, which are compatible with the upper bound $|\kappa_4| < 0.00024$ reported in \cite{Bazavov:2017dus}.

\begin{table}[h]
\begin{center}
\begin{tabular}{cccccc}
\hline\noalign{\smallskip}
 & $\chi^2_{phase}$ & $T_{0}$ (MeV) & $\mu_0$ (MeV) & a (GeV) & $\kappa_2$\\
\noalign{\smallskip}\hline\noalign{\smallskip}
ideal & 6.95 & 158.9$\pm$3.7 & 946$\pm$30 & 6.97$\pm$0.42 & 0.0181$\pm$0.0051\\
s-inv & 1.59 & 155.4$\pm$1.4 & 886$\pm$31 & 8.19$\pm$0.56 & 0.0139$\pm$0.0018\\
\noalign{\smallskip}\hline
\end{tabular}
\end{center}
\caption{Coefficients for the interpolations of the freeze-out parameters extracted with ideal and s-inv schemes employing Eqs.~(\ref{mub_energ}) and \ref{t_phase}, together with the reduced $\chi^2$ for Eq.~(\ref{t_phase}) ($\chi^2_{phase}$).}
\label{par-fo}
\end{table}

\section{Conclusions}

A systematic study of the effects of EV interactions, employing different schemes for the dependencies of particle eigenvolumes on mass and flavour, was performed. The difference between light (non-strange) and strange hadrons critically improves the description of ALICE hadron yield data in most central collisions at 2.76 TeV, the best agreement obtained in the s-inv scheme where the effective sizes of light unflavoured hadrons are proportional to their mass and the ones for strange hadrons are inversely proportional to their mass. This suggests that flavor-dependent eigenvolume interactions are a possible explanation of the so-called proton anomaly. 

Without any further tuning of the model parameters, the s-inv scheme is then used to investigate the thermal fits at different ALICE centralities and at different collision energies. A systematic improvement in the quality of the hadron yield data description over the ideal HRG model is obtained in all cases. 

The thermal fits within the s-inv scheme result in a modified chemical freeze-out curve [Eq. (\ref{t_phase})], with the extracted curvature being rather similar to some recent estimates of the curvature of the pseudo-critical QCD transition curve obtained on the lattice~\cite{Bellwied:2015rza,Bonati:2015bha}.

Only particles listed by the PDG are considered in the hadronic list here employed. In terms of data-model deviations, our analysis using this list shows a preference for the "exotic" s-inv scheme over other considered parameterisations. This result can possibly be understood as the absence of the assumingly existent, but as of yet unmeasured, strange baryons \cite{Bazavov:2014xya,Alba:2017mqu}. As shown in Ref. \cite{Alba:2017bbr}, once these unmeasured strange baryons are added to the particle list, the best description of lattice QCD simulations is achieved in the EV scheme where the hadronic sizes of all flavours are directly proportional to their mass.
Mass proportional hadron volumes is a rather expected trend, deriving from both radial and angular excitations, with a smaller volume for particles which contain the heavier strange quarks with respect to hadrons with approximately the same mass but made entirely of light quarks. Both these features are also in line with experimental measurements of the charged radii of the ground state hadrons~\cite{Patrignani:2016xqp}.
The investigations of the EV effects on chemical freeze-out conditions can be extended by employing a larger set of experimental data. These may include, e.g., the higher order moments of conserved charges fluctuations~\cite{Alba:2014eba}, which are currently being measured in experiments.
The inclusion of the extra strange baryons is another possibility that will be explored.

\section*{Acknowledgments}

We are thankful to D. Oliinychenko for fruitful discussions. This work was supported by HIC for FAIR within the LOEWE program of the State of Hesse.
H.St. acknowledges the support through the Judah M. Eisenberg Laureatus Chair at Goethe University. V.V. acknowledges the support from HGS-HIRe. The work of M.I.G. was supported by the Program of Fundamental Research of the Department of Physics and Astronomy of National Academy of Sciences of Ukraine.

\section*{References}

\bibliography{bib_flav_lhc}

\begin{thebibliography}{10}
\expandafter\ifx\csname url\endcsname\relax
  \def\url#1{\texttt{#1}}\fi
\expandafter\ifx\csname urlprefix\endcsname\relax\def\urlprefix{URL }\fi
\expandafter\ifx\csname href\endcsname\relax
  \def\href#1#2{#2} \def\path#1{#1}\fi

\bibitem{Stoecker:1981za}
H.~Stoecker, A.~A. Ogloblin, W.~Greiner, {SIGNIFICANCE OF TEMPERATURE
  MEASUREMENTS IN RELATIVISTIC NUCLEAR COLLISIONS}, Z. Phys. A303 (1981)
  259--266.
\newblock \href {http://dx.doi.org/10.1007/BF01421522}
  {\path{doi:10.1007/BF01421522}}.

\bibitem{Cleymans:1992zc}
J.~Cleymans, H.~Satz, {Thermal hadron production in high-energy heavy ion
  collisions}, Z. Phys. C57 (1993) 135--148.
\newblock \href {http://arxiv.org/abs/hep-ph/9207204}
  {\path{arXiv:hep-ph/9207204}}, \href {http://dx.doi.org/10.1007/BF01555746}
  {\path{doi:10.1007/BF01555746}}.

\bibitem{Yen:1998pa}
G.~D. Yen, M.~I. Gorenstein, {The Analysis of particle multiplicities in Pb +
  Pb collisions at CERN SPS within hadron gas models}, Phys. Rev. C59 (1999)
  2788--2791.
\newblock \href {http://arxiv.org/abs/nucl-th/9808012}
  {\path{arXiv:nucl-th/9808012}}, \href
  {http://dx.doi.org/10.1103/PhysRevC.59.2788}
  {\path{doi:10.1103/PhysRevC.59.2788}}.

\bibitem{Becattini:2000jw}
F.~Becattini, J.~Cleymans, A.~Keranen, E.~Suhonen, K.~Redlich, {Features of
  particle multiplicities and strangeness production in central heavy ion
  collisions between 1.7A-GeV/c and 158A-GeV/c}, Phys. Rev. C64 (2001) 024901.
\newblock \href {http://arxiv.org/abs/hep-ph/0002267}
  {\path{arXiv:hep-ph/0002267}}, \href
  {http://dx.doi.org/10.1103/PhysRevC.64.024901}
  {\path{doi:10.1103/PhysRevC.64.024901}}.

\bibitem{BraunMunzinger:2001ip}
P.~Braun-Munzinger, D.~Magestro, K.~Redlich, J.~Stachel, {Hadron production in
  Au - Au collisions at RHIC}, Phys. Lett. B518 (2001) 41--46.
\newblock \href {http://arxiv.org/abs/hep-ph/0105229}
  {\path{arXiv:hep-ph/0105229}}, \href
  {http://dx.doi.org/10.1016/S0370-2693(01)01069-3}
  {\path{doi:10.1016/S0370-2693(01)01069-3}}.

\bibitem{Rafelski:2002ga}
J.~Rafelski, J.~Letessier, {Testing limits of statistical hadronization}, Nucl.
  Phys. A715 (2003) 98--107.
\newblock \href {http://arxiv.org/abs/nucl-th/0209084}
  {\path{arXiv:nucl-th/0209084}}, \href
  {http://dx.doi.org/10.1016/S0375-9474(02)01418-5}
  {\path{doi:10.1016/S0375-9474(02)01418-5}}.

\bibitem{Andronic:2005yp}
A.~Andronic, P.~Braun-Munzinger, J.~Stachel, {Hadron production in central
  nucleus-nucleus collisions at chemical freeze-out}, Nucl. Phys. A772 (2006)
  167--199.
\newblock \href {http://arxiv.org/abs/nucl-th/0511071}
  {\path{arXiv:nucl-th/0511071}}, \href
  {http://dx.doi.org/10.1016/j.nuclphysa.2006.03.012}
  {\path{doi:10.1016/j.nuclphysa.2006.03.012}}.

\bibitem{Vovchenko:2015idt}
V.~Vovchenko, V.~V. Begun, M.~I. Gorenstein, {Hadron multiplicities and
  chemical freeze-out conditions in proton-proton and nucleus-nucleus
  collisions}, Phys. Rev. C93~(6) (2016) 064906.
\newblock \href {http://arxiv.org/abs/1512.08025} {\path{arXiv:1512.08025}},
  \href {http://dx.doi.org/10.1103/PhysRevC.93.064906}
  {\path{doi:10.1103/PhysRevC.93.064906}}.

\bibitem{Rischke:1991ke}
D.~H. Rischke, M.~I. Gorenstein, H.~Stoecker, W.~Greiner, {Excluded volume
  effect for the nuclear matter equation of state}, Z. Phys. C51 (1991)
  485--490.
\newblock \href {http://dx.doi.org/10.1007/BF01548574}
  {\path{doi:10.1007/BF01548574}}.

\bibitem{BraunMunzinger:1994xr}
P.~Braun-Munzinger, J.~Stachel, J.~P. Wessels, N.~Xu, {Thermal equilibration
  and expansion in nucleus-nucleus collisions at the AGS}, Phys. Lett. B344
  (1995) 43--48.
\newblock \href {http://arxiv.org/abs/nucl-th/9410026}
  {\path{arXiv:nucl-th/9410026}}, \href
  {http://dx.doi.org/10.1016/0370-2693(94)01534-J}
  {\path{doi:10.1016/0370-2693(94)01534-J}}.

\bibitem{Alba:2014eba}
P.~Alba, W.~Alberico, R.~Bellwied, M.~Bluhm, V.~Mantovani~Sarti, M.~Nahrgang,
  C.~Ratti, {Freeze-out conditions from net-proton and net-charge fluctuations
  at RHIC}, Phys. Lett. B738 (2014) 305--310.
\newblock \href {http://arxiv.org/abs/1403.4903} {\path{arXiv:1403.4903}},
  \href {http://dx.doi.org/10.1016/j.physletb.2014.09.052}
  {\path{doi:10.1016/j.physletb.2014.09.052}}.

\bibitem{Borsanyi:2013bia}
S.~Borsanyi, Z.~Fodor, C.~Hoelbling, S.~D. Katz, S.~Krieg, K.~K. Szabo, {Full
  result for the QCD equation of state with 2+1 flavors}, Phys. Lett. B730
  (2014) 99--104.
\newblock \href {http://arxiv.org/abs/1309.5258} {\path{arXiv:1309.5258}},
  \href {http://dx.doi.org/10.1016/j.physletb.2014.01.007}
  {\path{doi:10.1016/j.physletb.2014.01.007}}.

\bibitem{Bazavov:2014pvz}
A.~Bazavov, et~al., {Equation of state in ( 2+1 )-flavor QCD}, Phys. Rev. D90
  (2014) 094503.
\newblock \href {http://arxiv.org/abs/1407.6387} {\path{arXiv:1407.6387}},
  \href {http://dx.doi.org/10.1103/PhysRevD.90.094503}
  {\path{doi:10.1103/PhysRevD.90.094503}}.

\bibitem{Braun-Munzinger:2014lba}
P.~Braun-Munzinger, A.~Kalweit, K.~Redlich, J.~Stachel, {Confronting
  fluctuations of conserved charges in central nuclear collisions at the LHC
  with predictions from Lattice QCD}, Phys. Lett. B747 (2015) 292--298.
\newblock \href {http://arxiv.org/abs/1412.8614} {\path{arXiv:1412.8614}},
  \href {http://dx.doi.org/10.1016/j.physletb.2015.05.077}
  {\path{doi:10.1016/j.physletb.2015.05.077}}.

\bibitem{Andronic:2012ut}
A.~Andronic, P.~Braun-Munzinger, J.~Stachel, M.~Winn, {Interacting hadron
  resonance gas meets lattice QCD}, Phys. Lett. B718 (2012) 80--85.
\newblock \href {http://arxiv.org/abs/1201.0693} {\path{arXiv:1201.0693}},
  \href {http://dx.doi.org/10.1016/j.physletb.2012.10.001}
  {\path{doi:10.1016/j.physletb.2012.10.001}}.

\bibitem{Begun:2012rf}
V.~V. Begun, M.~Gazdzicki, M.~I. Gorenstein, {Hadron-resonance gas at
  freeze-out: Reminder on the importance of repulsive interactions}, Phys. Rev.
  C88~(2) (2013) 024902.
\newblock \href {http://arxiv.org/abs/1208.4107} {\path{arXiv:1208.4107}},
  \href {http://dx.doi.org/10.1103/PhysRevC.88.024902}
  {\path{doi:10.1103/PhysRevC.88.024902}}.

\bibitem{Vovchenko:2014pka}
V.~Vovchenko, D.~V. Anchishkin, M.~I. Gorenstein, {Hadron Resonance Gas
  Equation of State from Lattice QCD}, Phys. Rev. C91~(2) (2015) 024905.
\newblock \href {http://arxiv.org/abs/1412.5478} {\path{arXiv:1412.5478}},
  \href {http://dx.doi.org/10.1103/PhysRevC.91.024905}
  {\path{doi:10.1103/PhysRevC.91.024905}}.

\bibitem{Albright:2014gva}
M.~Albright, J.~Kapusta, C.~Young, {Matching Excluded Volume Hadron Resonance
  Gas Models and Perturbative QCD to Lattice Calculations}, Phys. Rev. C90~(2)
  (2014) 024915.
\newblock \href {http://arxiv.org/abs/1404.7540} {\path{arXiv:1404.7540}},
  \href {http://dx.doi.org/10.1103/PhysRevC.90.024915}
  {\path{doi:10.1103/PhysRevC.90.024915}}.

\bibitem{Alba:2017bbr}
P.~Alba, {The balance of attractive and repulsive hadronic interactions: the
  influence of hadronic spectrum and excluded volume effects on lattice
  thermodynamics and consequences on experiments}\href
  {http://arxiv.org/abs/1711.02797} {\path{arXiv:1711.02797}}.

\bibitem{Bugaev:2012wp}
K.~A. Bugaev, D.~R. Oliinychenko, A.~S. Sorin, G.~M. Zinovjev, {Simple Solution
  to the Strangeness Horn Description Puzzle}, Eur. Phys. J. A49 (2013) 30.
\newblock \href {http://arxiv.org/abs/1208.5968} {\path{arXiv:1208.5968}},
  \href {http://dx.doi.org/10.1140/epja/i2013-13030-y}
  {\path{doi:10.1140/epja/i2013-13030-y}}.

\bibitem{Vovchenko:2017zpj}
V.~Vovchenko, A.~Motornenko, P.~Alba, M.~I. Gorenstein, L.~M. Satarov,
  H.~Stoecker, {Multicomponent van der Waals equation of state: Applications in
  nuclear and hadronic physics}, Phys. Rev. C96~(4) (2017) 045202.
\newblock \href {http://arxiv.org/abs/1707.09215} {\path{arXiv:1707.09215}},
  \href {http://dx.doi.org/10.1103/PhysRevC.96.045202}
  {\path{doi:10.1103/PhysRevC.96.045202}}.

\bibitem{Vovchenko:2015cbk}
V.~Vovchenko, H.~Stšcker, {Surprisingly large uncertainties in temperature
  extraction from thermal fits to hadron yield data at LHC}, J. Phys. G44~(5)
  (2017) 055103.
\newblock \href {http://arxiv.org/abs/1512.08046} {\path{arXiv:1512.08046}},
  \href {http://dx.doi.org/10.1088/1361-6471/aa6409}
  {\path{doi:10.1088/1361-6471/aa6409}}.

\bibitem{Abelev:2013vea}
B.~Abelev, et~al., {Centrality dependence of $\pi$, K, p production in Pb-Pb
  collisions at $\sqrt{s_{NN}}$ = 2.76 TeV}, Phys. Rev. C88 (2013) 044910.
\newblock \href {http://arxiv.org/abs/1303.0737} {\path{arXiv:1303.0737}},
  \href {http://dx.doi.org/10.1103/PhysRevC.88.044910}
  {\path{doi:10.1103/PhysRevC.88.044910}}.

\bibitem{Abelev:2013xaa}
B.~B. Abelev, et~al., {$K^0_S$ and $\Lambda$ production in Pb-Pb collisions at
  $\sqrt{s_{NN}}$ = 2.76 TeV}, Phys. Rev. Lett. 111 (2013) 222301.
\newblock \href {http://arxiv.org/abs/1307.5530} {\path{arXiv:1307.5530}},
  \href {http://dx.doi.org/10.1103/PhysRevLett.111.222301}
  {\path{doi:10.1103/PhysRevLett.111.222301}}.

\bibitem{ABELEV:2013zaa}
B.~B. Abelev, et~al., {Multi-strange baryon production at mid-rapidity in Pb-Pb
  collisions at $\sqrt{s_{NN}}$ = 2.76 TeV}, Phys. Lett. B728 (2014) 216--227,
  [Erratum: Phys. Lett.B734,409(2014)].
\newblock \href {http://arxiv.org/abs/1307.5543} {\path{arXiv:1307.5543}},
  \href {http://dx.doi.org/10.1016/j.physletb.2014.05.052,
  10.1016/j.physletb.2013.11.048} {\path{doi:10.1016/j.physletb.2014.05.052,
  10.1016/j.physletb.2013.11.048}}.

\bibitem{Becattini:2014hla}
F.~Becattini, E.~Grossi, M.~Bleicher, J.~Steinheimer, R.~Stock, {Centrality
  dependence of hadronization and chemical freeze-out conditions in heavy ion
  collisions at $\sqrt s_{NN}$ = 2.76 TeV}, Phys. Rev. C90~(5) (2014) 054907.
\newblock \href {http://arxiv.org/abs/1405.0710} {\path{arXiv:1405.0710}},
  \href {http://dx.doi.org/10.1103/PhysRevC.90.054907}
  {\path{doi:10.1103/PhysRevC.90.054907}}.

\bibitem{Abelev:2014uua}
B.~B. Abelev, et~al., {$K^*(892)^0$ and $\phi(1020)$ production in Pb-Pb
  collisions at $\sqrt{s{NN}}$ = 2.76 TeV}, Phys. Rev. C91 (2015) 024609.
\newblock \href {http://arxiv.org/abs/1404.0495} {\path{arXiv:1404.0495}},
  \href {http://dx.doi.org/10.1103/PhysRevC.91.024609}
  {\path{doi:10.1103/PhysRevC.91.024609}}.

\bibitem{Patrignani:2016xqp}
C.~Patrignani, et~al., {Review of Particle Physics}, Chin. Phys. C40~(10)
  (2016) 100001.
\newblock \href {http://dx.doi.org/10.1088/1674-1137/40/10/100001}
  {\path{doi:10.1088/1674-1137/40/10/100001}}.

\bibitem{Nahrgang:2014fza}
M.~Nahrgang, M.~Bluhm, P.~Alba, R.~Bellwied, C.~Ratti, {Impact of resonance
  regeneration and decay on the net-proton fluctuations in a hadron resonance
  gas}, Eur. Phys. J. C75~(12) (2015) 573.
\newblock \href {http://arxiv.org/abs/1402.1238} {\path{arXiv:1402.1238}},
  \href {http://dx.doi.org/10.1140/epjc/s10052-015-3775-0}
  {\path{doi:10.1140/epjc/s10052-015-3775-0}}.

\bibitem{Bluhm:2014wha}
M.~Bluhm, P.~Alba, W.~Alberico, R.~Bellwied, V.~Mantovani~Sarti, M.~Nahrgang,
  C.~Ratti, {Determination of freeze-out conditions from fluctuation
  observables measured at RHIC}, Nucl. Phys. A931 (2014) 814--819.
\newblock \href {http://arxiv.org/abs/1408.4734} {\path{arXiv:1408.4734}},
  \href {http://dx.doi.org/10.1016/j.nuclphysa.2014.08.016}
  {\path{doi:10.1016/j.nuclphysa.2014.08.016}}.

\bibitem{Adam:2015vda}
J.~Adam, et~al., {Production of light nuclei and anti-nuclei in pp and Pb-Pb
  collisions at energies available at the CERN Large Hadron Collider}, Phys.
  Rev. C93~(2) (2016) 024917.
\newblock \href {http://arxiv.org/abs/1506.08951} {\path{arXiv:1506.08951}},
  \href {http://dx.doi.org/10.1103/PhysRevC.93.024917}
  {\path{doi:10.1103/PhysRevC.93.024917}}.

\bibitem{Adam:2015yta}
J.~Adam, et~al., {$^{3}_{\Lambda}\mathrm H$ and $^{3}_{\bar{\Lambda}}
  \overline{\mathrm H}$ production in Pb-Pb collisions at $\sqrt{s_{\rm NN}} =$
  2.76 TeV}, Phys. Lett. B754 (2016) 360--372.
\newblock \href {http://arxiv.org/abs/1506.08453} {\path{arXiv:1506.08453}},
  \href {http://dx.doi.org/10.1016/j.physletb.2016.01.040}
  {\path{doi:10.1016/j.physletb.2016.01.040}}.

\bibitem{Andronic:2017pug}
A.~Andronic, P.~Braun-Munzinger, K.~Redlich, J.~Stachel, {Decoding the phase
  structure of QCD via particle production at high energy}\href
  {http://arxiv.org/abs/1710.09425} {\path{arXiv:1710.09425}}.

\bibitem{Vovchenko:2016ebv}
V.~Vovchenko, H.~Stoecker, {Examination of the sensitivity of the thermal fits
  to heavy-ion hadron yield data to the modeling of the eigenvolume
  interactions}, Phys. Rev. C95~(4) (2017) 044904.
\newblock \href {http://arxiv.org/abs/1606.06218} {\path{arXiv:1606.06218}},
  \href {http://dx.doi.org/10.1103/PhysRevC.95.044904}
  {\path{doi:10.1103/PhysRevC.95.044904}}.

\bibitem{Friedmann:2009mx}
T.~Friedmann, {No Radial Excitations in Low Energy QCD. I. Diquarks and
  Classification of Mesons}, Eur. Phys. J. C73~(2) (2013) 2298.
\newblock \href {http://arxiv.org/abs/0910.2229} {\path{arXiv:0910.2229}},
  \href {http://dx.doi.org/10.1140/epjc/s10052-013-2298-9}
  {\path{doi:10.1140/epjc/s10052-013-2298-9}}.

\bibitem{Friedmann:2009mz}
T.~Friedmann, {No Radial Excitations in Low Energy QCD. II. The Shrinking
  Radius of Hadrons}, Eur. Phys. J. C73~(2) (2013) 2299.
\newblock \href {http://arxiv.org/abs/0910.2231} {\path{arXiv:0910.2231}},
  \href {http://dx.doi.org/10.1140/epjc/s10052-013-2299-8}
  {\path{doi:10.1140/epjc/s10052-013-2299-8}}.

\bibitem{Friedmann:2012kr}
T.~Friedmann, {A new QCD effect: The shrinking radius of hadrons}, Acta Phys.
  Polon. Supp. 5 (2012) 695--706.
\newblock \href {http://dx.doi.org/10.5506/APhysPolBSupp.5.695}
  {\path{doi:10.5506/APhysPolBSupp.5.695}}.

\bibitem{Alba:2016fku}
P.~Alba, W.~M. Alberico, A.~Nada, M.~Panero, H.~Stšcker, {Excluded-volume
  effects for a hadron gas in Yang-Mills theory}, Phys. Rev. D95~(9) (2017)
  094511.
\newblock \href {http://arxiv.org/abs/1611.05872} {\path{arXiv:1611.05872}},
  \href {http://dx.doi.org/10.1103/PhysRevD.95.094511}
  {\path{doi:10.1103/PhysRevD.95.094511}}.

\bibitem{Bazavov:2014xya}
A.~Bazavov, et~al., {Additional Strange Hadrons from QCD Thermodynamics and
  Strangeness Freezeout in Heavy Ion Collisions}, Phys. Rev. Lett. 113~(7)
  (2014) 072001.
\newblock \href {http://arxiv.org/abs/1404.6511} {\path{arXiv:1404.6511}},
  \href {http://dx.doi.org/10.1103/PhysRevLett.113.072001}
  {\path{doi:10.1103/PhysRevLett.113.072001}}.

\bibitem{Alba:2017mqu}
P.~Alba, et~al., {Constraining the hadronic spectrum through QCD thermodynamics
  on the lattice}, Phys. Rev. D96~(3) (2017) 034517.
\newblock \href {http://arxiv.org/abs/1702.01113} {\path{arXiv:1702.01113}},
  \href {http://dx.doi.org/10.1103/PhysRevD.96.034517}
  {\path{doi:10.1103/PhysRevD.96.034517}}.

\bibitem{Petran:2013lja}
M.~Petr‡n, J.~Letessier, V.~Petr‡cek, J.~Rafelski, {Hadron production and
  quark-gluon plasma hadronization in Pb-Pb collisions at $\sqrt{s_{NN}}=2.76$
  TeV}, Phys. Rev. C88~(3) (2013) 034907.
\newblock \href {http://arxiv.org/abs/1303.2098} {\path{arXiv:1303.2098}},
  \href {http://dx.doi.org/10.1103/PhysRevC.88.034907}
  {\path{doi:10.1103/PhysRevC.88.034907}}.

\bibitem{Stachel:2013zma}
J.~Stachel, A.~Andronic, P.~Braun-Munzinger, K.~Redlich, {Confronting LHC data
  with the statistical hadronization model}, J. Phys. Conf. Ser. 509 (2014)
  012019.
\newblock \href {http://arxiv.org/abs/1311.4662} {\path{arXiv:1311.4662}},
  \href {http://dx.doi.org/10.1088/1742-6596/509/1/012019}
  {\path{doi:10.1088/1742-6596/509/1/012019}}.

\bibitem{Floris:2014pta}
M.~Floris, {Hadron yields and the phase diagram of strongly interacting
  matter}, Nucl. Phys. A931 (2014) 103--112.
\newblock \href {http://arxiv.org/abs/1408.6403} {\path{arXiv:1408.6403}},
  \href {http://dx.doi.org/10.1016/j.nuclphysa.2014.09.002}
  {\path{doi:10.1016/j.nuclphysa.2014.09.002}}.

\bibitem{Steinheimer:2012rd}
J.~Steinheimer, J.~Aichelin, M.~Bleicher, {Nonthermal p/¹ Ratio at LHC as a
  Consequence of Hadronic Final State Interactions}, Phys. Rev. Lett. 110~(4)
  (2013) 042501.
\newblock \href {http://arxiv.org/abs/1203.5302} {\path{arXiv:1203.5302}},
  \href {http://dx.doi.org/10.1103/PhysRevLett.110.042501}
  {\path{doi:10.1103/PhysRevLett.110.042501}}.

\bibitem{Becattini:2012xb}
F.~Becattini, M.~Bleicher, T.~Kollegger, T.~Schuster, J.~Steinheimer, R.~Stock,
  {Hadron Formation in Relativistic Nuclear Collisions and the QCD Phase
  Diagram}, Phys. Rev. Lett. 111 (2013) 082302.
\newblock \href {http://arxiv.org/abs/1212.2431} {\path{arXiv:1212.2431}},
  \href {http://dx.doi.org/10.1103/PhysRevLett.111.082302}
  {\path{doi:10.1103/PhysRevLett.111.082302}}.

\bibitem{Adams:2003xp}
J.~Adams, et~al., {Identified particle distributions in pp and Au+Au collisions
  at s(NN)**(1/2) = 200 GeV}, Phys. Rev. Lett. 92 (2004) 112301.
\newblock \href {http://arxiv.org/abs/nucl-ex/0310004}
  {\path{arXiv:nucl-ex/0310004}}, \href
  {http://dx.doi.org/10.1103/PhysRevLett.92.112301}
  {\path{doi:10.1103/PhysRevLett.92.112301}}.

\bibitem{Adams:2006ke}
J.~Adams, et~al., {Scaling Properties of Hyperon Production in Au+Au Collisions
  at s**(1/2) = 200-GeV}, Phys. Rev. Lett. 98 (2007) 062301.
\newblock \href {http://arxiv.org/abs/nucl-ex/0606014}
  {\path{arXiv:nucl-ex/0606014}}, \href
  {http://dx.doi.org/10.1103/PhysRevLett.98.062301}
  {\path{doi:10.1103/PhysRevLett.98.062301}}.

\bibitem{Abelev:2007rw}
B.~I. Abelev, et~al., {Partonic flow and phi-meson production in Au + Au
  collisions at s(NN)**(1/2) = 200-GeV}, Phys. Rev. Lett. 99 (2007) 112301.
\newblock \href {http://arxiv.org/abs/nucl-ex/0703033}
  {\path{arXiv:nucl-ex/0703033}}, \href
  {http://dx.doi.org/10.1103/PhysRevLett.99.112301}
  {\path{doi:10.1103/PhysRevLett.99.112301}}.

\bibitem{Afanasiev:2002mx}
S.~V. Afanasiev, et~al., {Energy dependence of pion and kaon production in
  central Pb + Pb collisions}, Phys. Rev. C66 (2002) 054902.
\newblock \href {http://arxiv.org/abs/nucl-ex/0205002}
  {\path{arXiv:nucl-ex/0205002}}, \href
  {http://dx.doi.org/10.1103/PhysRevC.66.054902}
  {\path{doi:10.1103/PhysRevC.66.054902}}.

\bibitem{Alt:2006dk}
C.~Alt, et~al., {Energy and centrality dependence of anti-p and p production
  and the anti-Lambda/anti-p ratio in Pb+Pb collisions between 20/A-GeV and
  158/A-Gev}, Phys. Rev. C73 (2006) 044910.
\newblock \href {http://dx.doi.org/10.1103/PhysRevC.73.044910}
  {\path{doi:10.1103/PhysRevC.73.044910}}.

\bibitem{Alt:2007aa}
C.~Alt, et~al., {Pion and kaon production in central Pb + Pb collisions at 20-A
  and 30-A-GeV: Evidence for the onset of deconfinement}, Phys. Rev. C77 (2008)
  024903.
\newblock \href {http://arxiv.org/abs/0710.0118} {\path{arXiv:0710.0118}},
  \href {http://dx.doi.org/10.1103/PhysRevC.77.024903}
  {\path{doi:10.1103/PhysRevC.77.024903}}.

\bibitem{Alt:2008qm}
C.~Alt, et~al., {Energy dependence of Lambda and Xi production in central Pb+Pb
  collisions at A-20, A-30, A-40, A-80, and A-158 GeV measured at the CERN
  Super Proton Synchrotron}, Phys. Rev. C78 (2008) 034918.
\newblock \href {http://arxiv.org/abs/0804.3770} {\path{arXiv:0804.3770}},
  \href {http://dx.doi.org/10.1103/PhysRevC.78.034918}
  {\path{doi:10.1103/PhysRevC.78.034918}}.

\bibitem{Alt:2008iv}
C.~Alt, et~al., {Energy dependence of phi meson production in central Pb+Pb
  collisions at s(NN)**(1/2) = 6 to 17 GeV}, Phys. Rev. C78 (2008) 044907.
\newblock \href {http://arxiv.org/abs/0806.1937} {\path{arXiv:0806.1937}},
  \href {http://dx.doi.org/10.1103/PhysRevC.78.044907}
  {\path{doi:10.1103/PhysRevC.78.044907}}.

\bibitem{Alt:2004kq}
C.~Alt, et~al., {Omega- and anti-Omega+ production in central Pb + Pb
  collisions at 40-AGeV and 158-AGeV}, Phys. Rev. Lett. 94 (2005) 192301.
\newblock \href {http://arxiv.org/abs/nucl-ex/0409004}
  {\path{arXiv:nucl-ex/0409004}}, \href
  {http://dx.doi.org/10.1103/PhysRevLett.94.192301}
  {\path{doi:10.1103/PhysRevLett.94.192301}}.

\bibitem{Anticic:2011ny}
T.~Anticic, et~al., {Antideuteron and deuteron production in mid-central Pb+Pb
  collisions at 158$A$ GeV}, Phys. Rev. C85 (2012) 044913.
\newblock \href {http://arxiv.org/abs/1111.2588} {\path{arXiv:1111.2588}},
  \href {http://dx.doi.org/10.1103/PhysRevC.85.044913}
  {\path{doi:10.1103/PhysRevC.85.044913}}.

\bibitem{Friese:2002re}
V.~Friese, {Production of strange resonances in C + C and Pb + Pb collisions at
  158-A-GeV}, Nucl. Phys. A698 (2002) 487--490.
\newblock \href {http://dx.doi.org/10.1016/S0375-9474(01)01410-5}
  {\path{doi:10.1016/S0375-9474(01)01410-5}}.

\bibitem{Bazavov:2011nk}
A.~Bazavov, et~al., {The chiral and deconfinement aspects of the QCD
  transition}, Phys. Rev. D85 (2012) 054503.
\newblock \href {http://arxiv.org/abs/1111.1710} {\path{arXiv:1111.1710}},
  \href {http://dx.doi.org/10.1103/PhysRevD.85.054503}
  {\path{doi:10.1103/PhysRevD.85.054503}}.

\bibitem{Bellwied:2015rza}
R.~Bellwied, S.~Borsanyi, Z.~Fodor, J.~GŸnther, S.~D. Katz, C.~Ratti, K.~K.
  Szabo, {The QCD phase diagram from analytic continuation}, Phys. Lett. B751
  (2015) 559--564.
\newblock \href {http://arxiv.org/abs/1507.07510} {\path{arXiv:1507.07510}},
  \href {http://dx.doi.org/10.1016/j.physletb.2015.11.011}
  {\path{doi:10.1016/j.physletb.2015.11.011}}.

\bibitem{Bazavov:2017dus}
A.~Bazavov, et~al., {The QCD Equation of State to $\mathcal{O}(\mu_B^6)$ from
  Lattice QCD}, Phys. Rev. D95~(5) (2017) 054504.
\newblock \href {http://arxiv.org/abs/1701.04325} {\path{arXiv:1701.04325}},
  \href {http://dx.doi.org/10.1103/PhysRevD.95.054504}
  {\path{doi:10.1103/PhysRevD.95.054504}}.

\bibitem{Becattini:2016xct}
F.~Becattini, J.~Steinheimer, R.~Stock, M.~Bleicher, {Hadronization conditions
  in relativistic nuclear collisions and the QCD pseudo-critical line}, Phys.
  Lett. B764 (2017) 241--246.
\newblock \href {http://arxiv.org/abs/1605.09694} {\path{arXiv:1605.09694}},
  \href {http://dx.doi.org/10.1016/j.physletb.2016.11.033}
  {\path{doi:10.1016/j.physletb.2016.11.033}}.

\bibitem{Bonati:2015bha}
C.~Bonati, M.~D'Elia, M.~Mariti, M.~Mesiti, F.~Negro, F.~Sanfilippo, {Curvature
  of the chiral pseudocritical line in QCD: Continuum extrapolated results},
  Phys. Rev. D92~(5) (2015) 054503.
\newblock \href {http://arxiv.org/abs/1507.03571} {\path{arXiv:1507.03571}},
  \href {http://dx.doi.org/10.1103/PhysRevD.92.054503}
  {\path{doi:10.1103/PhysRevD.92.054503}}.

\end{thebibliography}

\end{document}